\documentclass[10pt,journal,compsoc]{IEEEtran}

\usepackage{amsmath}
\usepackage[ruled,linesnumbered]{algorithm2e}
\usepackage{soul}

\usepackage{comment}
\usepackage{textcomp}
\usepackage[keeplastbox]{flushend}

\usepackage{comment}

\newcommand{\Remove}[1]{{}}

\newcommand{\squishlist}{
   \begin{list}{$\bullet$}
    { \setlength{\itemsep}{0pt}      \setlength{\parsep}{0pt}
      \setlength{\topsep}{3pt}       \setlength{\partopsep}{0pt}
      \setlength{\listparindent}{-2pt}
      \setlength{\itemindent}{-5pt}
      \setlength{\leftmargin}{1em} \setlength{\labelwidth}{0em}
      \setlength{\labelsep}{0.5em} } }

\newcommand{\squishend}{
    \end{list}  }
    
\usepackage{multirow}  
\usepackage{comment}
\usepackage[ruled,linesnumbered]{algorithm2e}
\usepackage{listings}
\usepackage{xcolor}
\usepackage{footnote}
\usepackage{threeparttable}
\usepackage{paralist}
\usepackage{booktabs}

\usepackage{xcolor}
\usepackage[normalem]{ulem} 
\usepackage{tikz}
\usetikzlibrary{calc}

\newcommand*\blackcircled[1]{\tikz[baseline=(char.base)]{
        \node[shape=circle,fill={rgb,255:red,0;green,0;blue,0}, text=white, font=\small, inner sep=0.6pt] (char) {#1};}}

\usepackage{multirow}

\usepackage{framed}   

\usepackage{pifont}

\usepackage[ruled,linesnumbered]{algorithm2e}
\usepackage{xcolor}
\usepackage{booktabs}
\usepackage{graphicx}  
\usepackage{wrapfig}
\usepackage{subfigure}

\usepackage{fancyhdr}

\ifCLASSOPTIONcompsoc
  \usepackage[nocompress]{cite}
\else
  \usepackage{cite}
\fi

\ifCLASSINFOpdf
\else
\fi

\begin{document}

\title{Characterizing and Understanding \\Distributed GNN Training on GPUs}

\author{Haiyang~Lin,
        ~Mingyu~Yan,
        ~Xiaocheng~Yang,
        ~Mo~Zou,
        ~Wenming~Li,
        ~Xiaochun~Ye,
        ~Dongrui~Fan
\IEEEcompsocitemizethanks{\IEEEcompsocthanksitem H. Lin, M. Yan, M. Zou, W. Li, X. Ye and D. Fan are with SKLCA, Institute of Computing Technology (ICT), Chinese Academy of Sciences (CAS), Beijing 100864, China and also with the University of Chinese Academy of Sciences (UCAS), Beijing 100049, China. E-mail: $\{$linhaiyang18z, Yanmingyu, zoumo, liwenming, yexiaochun, fandr\}@ict.ac.cn

\IEEEcompsocthanksitem X. Yang is with Institute of Computing Technology (ICT), Chinese Academy of Sciences (CAS), Beijing 100864, China. E-mail: yangxiaocheng@ict.ac.cn}
\thanks{
This work was supported by the Strategic Priority Research Program of Chinese Academy of Sciences (Grant No. XDC05000000), National Natural Science Foundation of China (Grant No. 61732018 and 61872335), Austrian-Chinese Cooperative R\&D Project (FFG and CAS) (Grant No. 171111KYSB20200002), CAS Project for Young Scientists in Basic Research (Grant No. YSBR-029), and CAS Project for Youth Innovation Promotion Association. The corresponding author is Mingyu Yan.
}
}

\markboth{
}%
{Shell \MakeLowercase{\textit{et al.}}: Bare Demo of IEEEtran.cls for Computer Society Journals}

\IEEEtitleabstractindextext{%
\begin{abstract}
Graph neural network (GNN) has been demonstrated to be a powerful model in many domains for its effectiveness in learning over graphs. 
To scale GNN training for large graphs, a widely adopted approach is distributed training which accelerates training using multiple computing nodes.
Maximizing the performance is essential, but the execution of distributed GNN training remains preliminarily understood.
In this work, we provide an in-depth analysis of distributed GNN training on GPUs, revealing several significant observations and providing useful guidelines for both software optimization and hardware optimization.
\end{abstract}

}

\maketitle

\IEEEdisplaynontitleabstractindextext

\IEEEpeerreviewmaketitle

\IEEEraisesectionheading{
\section{Introduction}\label{sec:introduction}}
Graph neural networks (GNNs) have achieved huge success in handling graph data, and have been applied to various applications in social networks, recommendations, etc. 
Scaling GNN training for large graphs, which consists of millions or even billions of edges \cite{dataset_OGB}, is exposed to significant time overhead.
In order to shorten the deployment time of GNN, distributed training is a widely adopted method that equips the training system with more computing resources to accelerate the training of GNN \cite{DistDGL, PaGraph, DistGNN}.

\thispagestyle{fancy}         
\fancyhead{}                     
\lhead{}          
\chead{\ifthenelse{\value{page}=1}{To Appear in IEEE Computer Architecture Letters (CAL) 2022}{}}
\rhead{}
\lfoot{}
\cfoot{}  
\rfoot{}

Distributed training of GNNs can be approached in two ways - distributed full-batch training \cite{DistGNN} and distributed mini-batch training \cite{DistDGL, PaGraph}.
Distributed full-batch training updates the GNN model with the whole graph in every forward and backward propagation \cite{DistGNN}, while distributed mini-batch training only uses part of vertices and edges through sampling method \cite{DistDGL, PaGraph}.
Distributed mini-batch training is more efficient than distributed full-batch training as it needs much less time to converge on large graphs while maintaining accuracy \cite{DistDGLv2}.
In this work, we focus on distributed mini-batch training on GPUs.

Many researchers are working on optimizing GNN training while the execution of it remains preliminarily understood, resulting in the inefficiency of optimization \cite{yan_gcn}.
To systematically understand GNN training on GPUs, analysis of the end-to-end execution is urgently needed.
However, the existing characterization work on GNN training execution focuses on single GPU execution \cite{yan_gcn, yan2020hygcn}, while the end-to-end execution of distributed GNN training has not been systematically analyzed, especially from the perspective of the parallel workers' interaction.
To the best of our knowledge, this is the first characterization work on the end-to-end execution of distributed GNN training.

In this work, we have an in-depth analysis of distributed GNN training through profiling the end-to-end execution with the state-of-the-art framework, Pytorch-Geometric (PyG) \cite{framework_PyG}, revealing several significant observations and providing useful guidelines for both software optimization and hardware optimization. 
To have a comprehensive knowledge of the end-to-end execution, we divide the execution into four phases - \emph{Sampling, Data Loading, Computing (Forward and Backward Propagation), Gradient Synchronization}.
We first provide an overview of the end-to-end execution and then probe into the interaction of the parallel workers in each phase to find out the factors hindering the performance improvement.
The key observations and insights are summarized below:

\begin{itemize}
    \item \emph{Overview of the End-to-End Execution:} \blackcircled{1} The data loading phase is the most time-consuming phase. 
    \blackcircled{2} The gradient synchronization phase becomes more time-consuming when the number of GPUs increases. 
    \blackcircled{3} The acceleration ratio using multiple GPUs has a significant gap compared to the ideal acceleration ratio.
    \item \emph{Interaction of the Parallel Workers:} \blackcircled{4} Serious competition of CPU shared cache between the sampling threads. 
    \blackcircled{5} Non-negligible bandwidth competition during the data loading phase.
    \blackcircled{6} Increasing synchronization time due to imbalance of CPU side.
\end{itemize}

Base on these insights, we provide several useful guidelines for both software and hardware optimizations. 
\begin{itemize}
    \item \emph{Software Optimizations:} 
    1) Localized Sampling for \blackcircled{4} - use clustering algorithm for vertices to improve cache reuse. 
    2) Pipeline Overlap and Caching for \blackcircled{5} - pipeline the data transferring and computing, and cache frequently used features in GPU memory. 
    3) Workload Balance Strategy for \blackcircled{6} - equip the system with backup threads to alleviate the imbalance effect.
    \item \emph{Hardware Optimizations:}
    1) Separated Cache for \blackcircled{4} - enable the shared cache to support separate cache management. 
    2) Hybrid Architecture for \blackcircled{5} - conduct sampling and computing on hybrid architecture. 
    3) Data Compression for \blackcircled{5} - use compression to minimize the size of transmitted data.
\end{itemize}

\vspace{-19pt}
\section{Background and Related Work}\label{BG}
\vspace{-3pt}

\textbf{Graph Neural Network:}
GNN has been dominated to be an effective model for learning knowledge from structured graph data. 
A GNN model consists of one or multiple layers and each GNN layer is composed of neighbor aggregation and neural network operations. 
The neighbor aggregation means gathering the activations of the neighbor vertices from the previous GNN layer while the neural network operations are applied to update the activations of the vertices. 
The computation is formally expressed as follow:
\begin{equation}\label{eq1}
    s_v^i = Aggregate(\{h_u^{i-1}|u \in N(v)\})
\end{equation}
\begin{equation}\label{eq2}
    h_v^i = Update(s_v^i, h_v^{i-1})
\end{equation}
where $h_v^i$ is the activation of vertex $v$ at the $i$-th layer and $N(v)$ donates the neighbors of vertex $v$.
$s_v^i$ is the aggregation result of vertex $v$ at the $i$-th layer.
$Aggregate(~)$ and $Update(~)$ are customized or parameterized functions.

\begin{figure}[hbtp]
    \vspace{-15pt}
    \centering
    \includegraphics[page=1, width=0.42\textwidth]{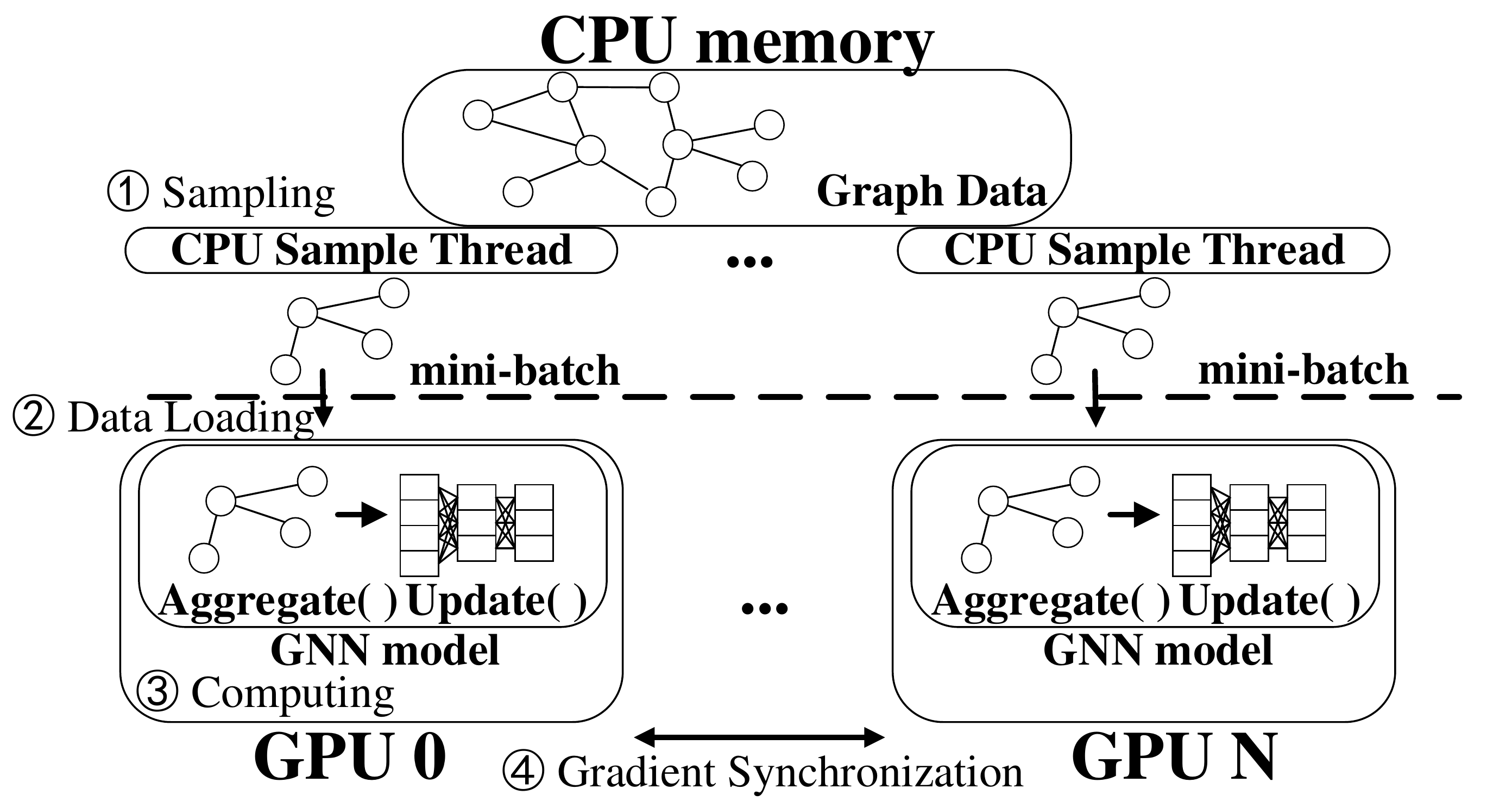}
    \vspace{-15pt}
    \caption{Mini-batch distributed GNN training.}
    \label{fig:01dt_gnn}
    \vspace{-20pt}
\end{figure}

\textbf{Mini-batch Distributed GNN Training:}
mini-batch training is more and more popular nowadays in distributed training of GNN \cite{DistDGL, PaGraph}.
Fig. \ref{fig:01dt_gnn} demonstrates the execution of mini-batch distributed GNN training which can be divided into four phases - \emph{Sampling, Data Loading, Computing (Forward and Backward Propagation), Gradient Synchronization}.
The whole graph data is kept on the host side, i.e., CPU memory, as the scale of data is too large for GPU memory. 
In each epoch, the CPU threads sample mini-batches consisting of partial vertices and edges \textcircled{1}. 
The mini-batches are then loaded to GPU memory for forward propagation and backward propagation \textcircled{2}, \textcircled{3}.
Then GPUs communicate with each other for synchronization to update the weights of the GNN model \textcircled{4}.

\vspace{-7pt}
\section{Evaluation Setup}\label{ES}
\vspace{-3pt}

\textbf{GNNs and Datasets:} 
we choose three popular GNN models - GCN \cite{model_gcn}, GAT \cite{model_GAT}, GraphSAGE \cite{model_GraphSAGE}, and three datasets - Reddit \cite{model_GraphSAGE}, Products \cite{dataset_OGB}, Papers \cite{dataset_OGB}.
Table \ref{table:dataset} lists the details of the three datasets.
The “feat" column represents the dimension of
vertex features and the “class" column represents the number
of vertex classes. 
Each GNN is composed of 2 layers and the mini-batch size is 1024 targe nodes. 
The hidden layer dimension is 256.
The sampling strategy is node-wise sampling \cite{model_GraphSAGE}.
It samples 10 neighbors in the first layer and 25 neighbors in the second layer.

\begin{table}[hbtp]
    \vspace{-12pt}
 \caption{Statistics of the datasets. \cite{model_GraphSAGE,dataset_OGB}} \label{table:dataset}
 \vspace{-10pt}
 \centering
 \renewcommand\arraystretch{1.0}
    \resizebox{0.45\textwidth}{!}{
\begin{tabular}{|c|r|r|r|r|}
\hline
\textbf{Datasets}    & \textbf{\#vertex}    & \textbf{\#edge} & \textbf{\#feat}  & \textbf{\#class}\\ \hline \hline
             Reddit             &232,965       &114,615,892   &602  &41   \\
             Products      &2,449,029     &123,718,280   &100  &47   \\
             Papers        &111,059,956   &1,615,685,872 &128  &172  \\
			 
\hline 
\end{tabular}
    }
    \vspace{-8pt}
\end{table}

\textbf{Profiling Platform:} 
we deploy the experiments on a multi-GPU server which consists of 2 Intel Xeon Silver 4208 CPUs (each with 8 cores) and 4 NVIDIA Tesla T4 GPUs. 
Each socket has two GPUs. 
The communication library is NCCL v2.10.3 and the gradient synchronization uses All-Reduce strategy.
The GPUs use PCIe 3.0 x8.
The machine is installed with Ubuntu 20.04.3 LTS, CUDA library v11.3, PyTorch v1.10, and PyG v2.0.2 \cite{framework_PyG}. 
We use NVIDIA Nsight Systems v2021.2.1 to analyze the details in GPU execution. 
We use Perf to obtain the CPU execution performance indexes.
We set the number of PyTorch dataloader workers for each process as 4 to fully utilize the CPU resource as the number of CPU cores is 16.
The class of dataloader we used is PyG NeighborLoader with default settings except for num\_workers and batch\_size.
The experimental data averages from 5 epochs execution.
We choose PyG, one of the most popular frameworks in GNN community.

\vspace{-17pt}
\section{Overview of the End-to-End Execution}\label{OE}
\vspace{-3pt}

In this section, we provide an overview of the end-to-end execution toward distributed GNN training.

\begin{figure*}[!hbtp]
    \vspace{-10pt}
    \centering
    \includegraphics[width=0.96\textwidth]{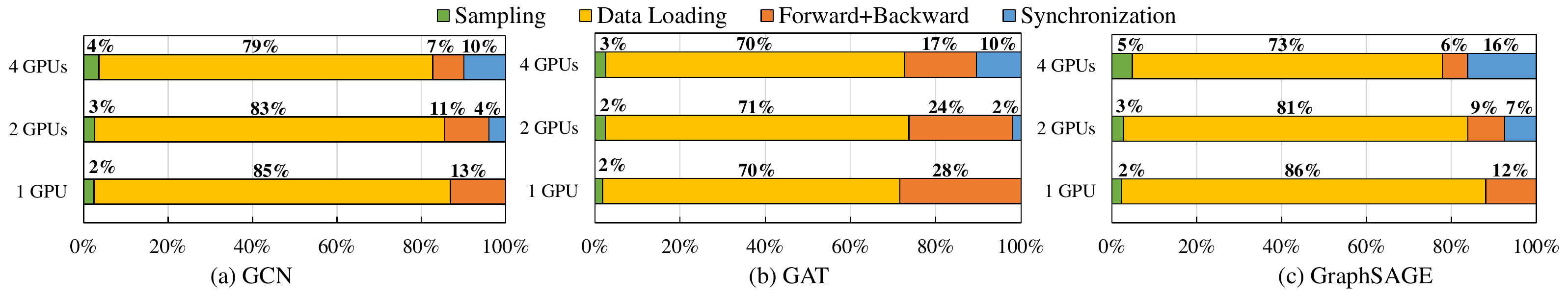}
    \vspace{-15pt}
    \caption{Execution time breakdown for three GNNs on 1,2,4 GPUs. The dataset is Reddit. (a) GCN (b) GAT (c) GraphSAGE.}
    \label{fig:ex01}
    \vspace{-16pt}
\end{figure*}

\textbf{Observation \blackcircled{1}:} \emph{The data loading phase is the most time-consuming phase.}

Fig. \ref{fig:ex01} illustrates the execution time breakdown for GCN, GAT, GraphSAGE on 1, 2, 4 GPUs. 
The data loading phase is the most time-consuming phase for all three GNNs.
For GCN, it occupies a proportion of 79\% to 85\% in one epoch. 
And the number is 70\% to 71\% for GAT and 73\% to 86\% for GraphSAGE.
When more GPUs are involved in distributed GNN training, the proportion of the data loading phase is becoming smaller, but still is the largest.
As for the sampling phase, it occupies only a small proportion of the execution time which is no more than 5\%. 
And the proportion of the computing phase, which is composed of forward propagation and backward propagation, is becoming smaller when the number of GPUs increases.

\textbf{Observation \blackcircled{2}:} \emph{The gradient synchronization phase becomes more time-consuming when the number of GPUs increases.} 

The proportion of the gradient synchronization phase is becoming larger dramatically when more GPUs are involved. 
As shown in Fig. \ref{fig:ex01}, for GCN, the proportion reaches 10\% of the execution time on 4 GPUs, while the number is only 4\% on 2 GPUs.
As for GAT and GraphSAGE, the proportion on 4 GPUs is 10\% and 16\% respectively.
The gradient synchronization phase replaces the computing phase as the second largest phase in 4 GPUs execution.
It is reasonable to be convinced that the Gradient Synchronization phase is becoming a non-negligible component with more GPUs' participation in training. 

\begin{figure}[!hbtp]
    \vspace{-6pt}
    \centering
    \includegraphics[width=0.42\textwidth]{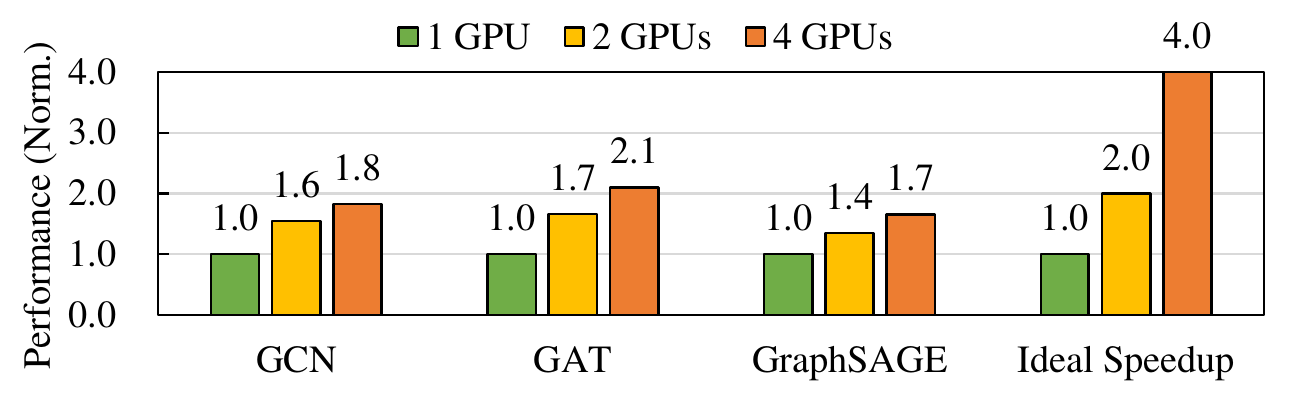}
    \vspace{-15pt}
    \caption{Normalized performance of distributed training using multiple GPUs compared to training on one GPU.}
    \label{fig:ex02}
    \vspace{-7pt}
\end{figure}

\textbf{Observation \blackcircled{3}:} \emph{The acceleration ratio using multiple GPUs has a significant gap compared to the ideal acceleration ratio.} 

Fig. \ref{fig:ex02} demonstrates the normalized performance of distributed training using multiple GPUs compared to training on one GPU.
Compared to 1 GPU execution, the acceleration ratio of 2 GPUs execution is 1.6$\times$, 1.7$\times$, and 1.4$\times$ for GCN, GAT, and GraphSAGE respectively while the ideal acceleration ratio is 2$\times$.
As for 4 GPUs execution, the acceleration ratio is 1.8$\times$, 2.1$\times$, 1.7$\times$ while the ideal acceleration ratio is 4$\times$.
The gap compared to the ideal acceleration ratio is not only significant but is becoming larger when the number of GPUs increases.
Other researches on this execution scene \cite{ispass01, ispass02_gnnmark} yield similar speedup results to ours, indicating that there are factors limiting the system's scalability.
However, they do not probe into the execution to find the reason. 
The execution pattern of mini-batch training on multi-GPU platform leads to frequent data movements between CPU and GPU, and resource competition between CPU threads, hindering the scalability of mini-batch distributed training.
In contrast, on multi-CPU platform, DistDGL \cite{DistDGL} scales almost linearly up to 16 nodes. 
Unlike multi-GPU execution, they perform sampling and GNN computing phases on the same CPU without abundant data movements. 
However, compared with GPU, the execution on CPU is much slower due to limited computing resources.
So multi-GPU is more common nowadays in distributed GNN training \cite{PaGraph,ispass01, ispass02_gnnmark}.

\begin{figure*}[!hbtp]
    \vspace{-10pt}
    \centering
    \includegraphics[width=0.96\textwidth]{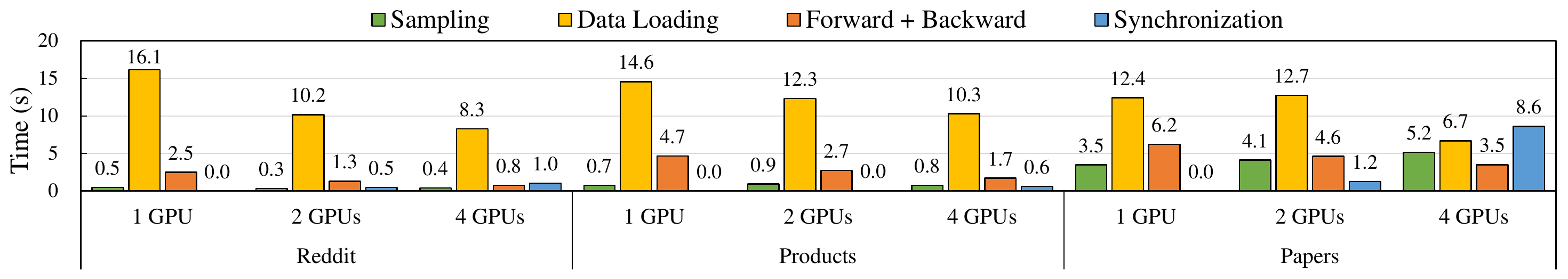}
    \vspace{-15pt}
    \caption{The execution time of each phase for GCN on 1,2,4 GPUs. The datasets are Reddit, Products and Papers.}
    \label{fig:ex03}
    \vspace{-10pt}
\end{figure*}

\vspace{-16pt}
\section{In-depth Analysis of the Interaction}\label{IA}
\vspace{-4pt}
In this section, we provide an in-depth analysis of the interaction of parallel workers in distributed GNN training.
Fig. \ref{fig:ex03} illustrates the execution time of each phase for GCN on 1, 2, and 4 GPUs. 
As the workloads of one epoch, i.e., target nodes, are distributed equally to parallel workers, it's expected that the execution time of each phase should decline when more GPUs are involved.
The time of the computing phase declines as expected, while the other three phases do not.
The reason easily comes to mind that only the computing phase of each worker is completely independent of others while the other three are not. 
So in the following, we probe into the other three phases to find out the factors hindering the performance improvement of distributed GNN training.

\begin{figure*}[!htbp]
    \centering
    \begin{minipage}{0.50\textwidth}
    \vspace{-2pt}
    \centering
    \includegraphics[width=\textwidth]{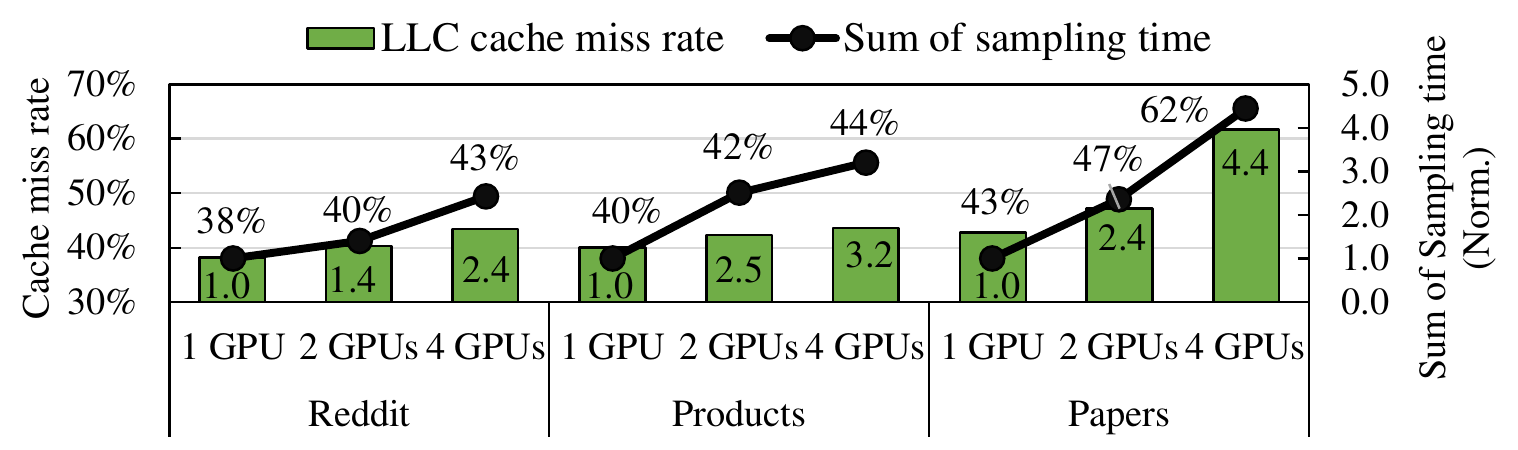}
    \vspace{-22pt}
    \caption{The Last Level Cache (LLC) miss rate and the sum of all sampling threads' sampling time, which is normalized and is used to reflect the real CPU execution time for sampling. The model is GCN.}
    \label{fig:ex04}
    \end{minipage}\hfill
    \begin{minipage}{0.47\textwidth}
    \vspace{-2pt}
    \centering
    \includegraphics[width=\textwidth]{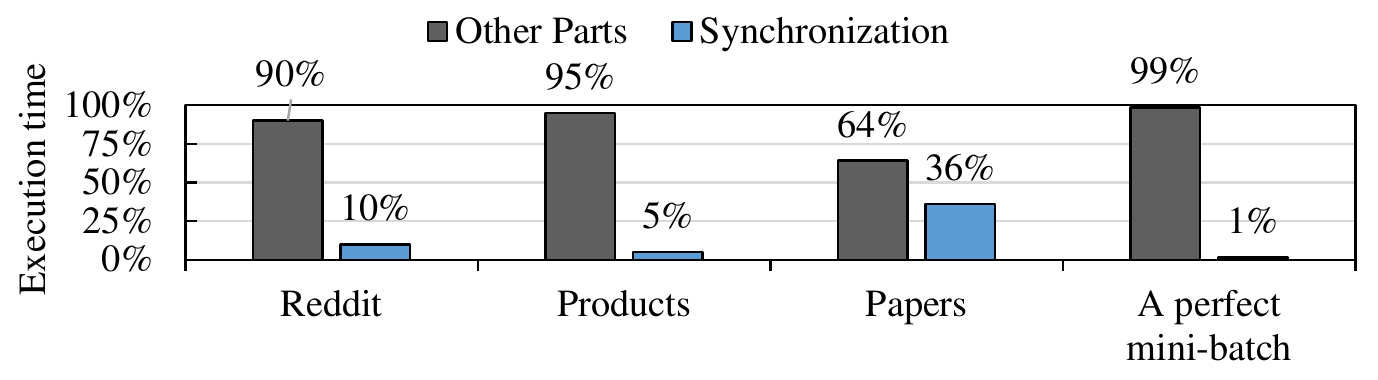}
    \vspace{-22pt}
    \caption{The execution time proportion of the gradient synchronization phase for GCN on different datasets using 4 GPUs. \emph{A perfect mini-batch} is a mini-batch in which four GPUs start computing phase almost at the same time.}
    \label{fig:ex05}
    \end{minipage}\hfill
    \vspace{-20pt}
\end{figure*}

\textbf{Observation \blackcircled{4}:} \emph{Serious competition of CPU shared cache between the sampling threads.} \textbf{(Sampling Phase)}

As shown in Fig. \ref{fig:ex03}, the time of the sampling phase does not decline sharply as expected when more GPUs are involved.
What's more, on the Papers dataset, the sampling time is even increasing, from 3.5$s$ to 5.2$s$ per epoch.
The sampling threads execute independently in CPU cores and compete for the CPU Last Level Cache (LLC).
Fig. \ref{fig:ex04} demonstrates the LLC miss rate and the sum of all sampling threads' sampling time, which is normalized and is used to reflect the real CPU execution time for sampling.
As more sampling threads are competing for the LLC cache, the LLC miss rate increases.
As a result, the sum of sampling time increases too.
The LLC miss rate increases from 38\% to 43\% for Reddit and from 40\% to 44\% for Products while the sum of sampling time for them reaches 2.4$\times$ and 3.2$\times$ respectively compared to 1 GPU execution.
As for Papers, the LLC miss rate increases dramatically from 43\% to 62\%.
Correspondingly, the sum of sampling time of 4 GPUs execution is 4.4$\times$ compared to 1 GPU execution. 
It validates the serious competition of CPU shared cache between the sampling threads, which hinders the performance improvement of distributed GNN training.

\textbf{Observation \blackcircled{5}:} \emph{
Non-negligible bandwidth competition during the data loading phase.} \textbf{(Data Loading Phase)}

Table \ref{table:bandwidth} illustrates the transmission bandwidth of Host-to-Device which means from CPU memory to GPU memory during the data loading phase. 
The model is GCN and the dataset is Reddit.
As the GPUs share PCIe channels to CPU with each other, it results in bandwidth decline when the number of GPUs participating in training increases.
The transfer bandwidth of 1 GPU training is 5.85 GiB/s while it drops to 3.68 GiB/s of 4 GPU training for each GPU.
It validates that PCIe competition between multiple GPUs has a negative impact on transmission bandwidth, thus hindering the performance improvement.
Moreover, we conduct the experiment with 1, 2, and 4 PyTorch dataloader workers for each process.
The execution time ratios of sampling and data loading phase are similar among different dataloader setups, indicating that the data loading overhead is not sensitive to dataloader worker numbers.

\begin{table}[hbtp]
    \vspace{-13pt}
 \caption{Data loading phase Host-to-Device transfer bandwidth.
 } \label{table:bandwidth}
 \vspace{-10pt}
 \centering
 \renewcommand\arraystretch{1.0}
    \resizebox{0.40\textwidth}{!}{
\begin{tabular}{|c|r|r|r|}
\hline
\textbf{~}    & \textbf{1 GPU} & \textbf{2 GPU} & \textbf{4 GPU}\\ \hline \hline
            Bandwidth (GiB/s)     &5.85    &4.75   &3.68 \\
\hline 
\end{tabular}
    }
    \vspace{-11pt}
\end{table}

\textbf{Observation \blackcircled{6}:} \emph{Increasing synchronization time due to imbalance of CPU side.} \textbf{(Gradient Synchronization Phase)}

Different from Deep Neural Network (DNN), the model size of GNN is always much smaller as it has few layers and shares weights across all vertices. 
It's expected that the gradient synchronization phase should only occupy a small part of the execution time \cite{PaGraph, DistDGL}.
However, as shown in Fig. \ref{fig:ex01}, the gradient synchronization phase takes a larger proportion of the execution time when the number of GPUs increases. 
Besides, Fig. \ref{fig:ex05} illustrates the execution time proportion of the gradient synchronization phase for GCN on different datasets using 4 GPUs.
The gradient synchronization phase takes up 10\% for Reddit, 5\% for Products, and 36\% for Papers.
We use the NVIDIA Nsight Systems to look into the execution and present a perfect mini-batch execution in Fig. \ref{fig:ex05}, in which four GPUs start computing phase almost at the same time. 
The gradient synchronization phase only takes up 1\% of the execution time in the perfect mini-batch, which indicates that the problem should lie in the imbalance of the CPU side.

The imbalance of the CPU side results in different start computing time of GPUs.
Table \ref{table:syn} presents the difference of GPU start computing time and the result is normalized to the execution time of an epoch.
The start computing time means the time when the GPU start computing phase.
The \emph{sum(Max-Min)} means summing up the difference of start computing time between the slowest and fastest GPU at each mini-batch.
And the \emph{sum(Max-Ave)} means summing up the difference of the slowest and the average start computing time of all GPUs.
The \emph{sum(Max-Min)} is 34\%, 17\%, 61\% for Reddit, Products, Papers respectively.
The \emph{sum(Max-Ave)} is more suitable to present the imbalance of the sampling and data loading phase.
It's 15\%, 8\%, 32\% for Reddit, Products, Papers and is close to the proportion of the gradient synchronization phase.
It validates that the \emph{stragglers} which means workers with abnormal performance also exist in the distributed GNN training as in DNN.

\begin{table}[hbtp]
    \vspace{-14pt}
 \caption{Imbalance of GPU start computing time. 
 } \label{table:syn}
 \vspace{-10pt}
 \centering
 \renewcommand\arraystretch{1.0}
    \resizebox{0.36\textwidth}{!}{
\begin{tabular}{|c|r|r|r|}
\hline
\textbf{~}    & \textbf{Reddit}    & \textbf{Products} & \textbf{Papers}\\ \hline \hline
             \emph{sum(Max-Min)}  &34\% &17\% &61\% \\
             \emph{sum(Max-Ave)}  &15\% &8\% &32\% \\
			 
\hline 
\end{tabular}
    }
    \vspace{-22pt}
\end{table}

\section{Architectural Guidelines}\label{AG}
\vspace{-5pt}

\textbf{Software Optimization Guidelines:}

\emph{Localized Sampling for \blackcircled{4}:} It means using clustering algorithms to preprocess the datasets and then distributing the target nodes in the same clusters to sampling threads in the same time period. In this way, the competition of LLC is alleviated as the target nodes of different sampling threads have plenty of common neighbors in the same time period.   

\emph{Pipeline Overlap and Caching for \blackcircled{5}:} The data loading phase is the major part of the execution time and it's independent of the computing in GPUs. It indicates that we can use the pipeline strategy to overlap the data transferring and computing. 
Another strategy is to cache features of vertices that are frequently visited in GPU memory to reduce the amount of data transferred.

\emph{Workload Balance Strategy for \blackcircled{6}:} As the imbalance hinders the performance, it's reasonable to use workload balance strategy, e.g., equipping the system with backup threads, to alleviate the impact of imbalance on performance.

\textbf{Hardware Optimization Guidelines:}

\emph{Separated Cache for \blackcircled{4}:} As the sampling threads compete for the LLC resource, they may corrupt the cache leading to frequent cache replacement. It may be a good choice to modify the cache to support separate cache strategy, which means each thread holds an independent part of the cache.

\emph{Hybrid Architecture for \blackcircled{5}:} The reason why sampling and computing are executed separately is that sampling is memory-intensive and computing is compute-intensive. It indicates that a hybrid architecture that supports both kinds of execution can eliminate the data loading phase.

\emph{Data Compression for \blackcircled{5}:} To shorten the overhead of data transferring, using data compression and decompression in the CPU and GPU side respectively is a good choice while it required specialized compress techniques with hardware support to minimize the overhead.

\vspace{-5pt}
\section{Conclusion}\label{Con}
\vspace{-3pt}
To scale GNN training for large graphs, distributed training is widely adopted, which uses multiple computing nodes to accelerate training.
However, the execution of distributed GNN training remains preliminarily understood, which brings difficulties to optimization. 
In this work, an in-depth analysis is carried out through profiling the execution of mini-batch distributed GNN training on GPUs, revealing several significant observations and providing optimization guidelines based on the observations.

\ifCLASSOPTIONcaptionsoff
  \newpage
\fi

\vspace{-17pt}
\bibliographystyle{IEEEtran}
\bibliography{arxiv}

\end{document}